\documentclass[aps,prb,reprint,groupedaddress,floatfix,a4paper]{revtex4-2}

\usepackage[english]{babel} 
\usepackage{amssymb,amsmath}
\usepackage{graphicx}
\usepackage{url}
\usepackage{multirow}
\usepackage{chemformula}
\usepackage{mathtools}
\usepackage{bm}
\usepackage{tabularx}
\usepackage{makecell}
\usepackage{xcolor}
\usepackage{siunitx}
\usepackage{comment}
\usepackage{hyperref}

\usepackage{subcaption}

\newlength{\smallpic}
\newcommand{\bmx}{\bm{x}}

\begin{document}

\title{Bayesian optimization of atomic structures with prior probabilities from universal interatomic potentials}
\author{Peder Lyngby$^*$, Casper Larsen, and Karsten Wedel Jacobsen
 }
 \address{Computational Atomic-scale Materials Design (CAMD), Department of Physics, Technical University of Denmark, 2800 Kgs. Lyngby Denmark \\ \normalfont{$^*$Corresponding author: pmely@dtu.dk}}
\date{\today}

\begin{abstract}
The optimization of atomic structures plays a pivotal role in understanding and designing materials with desired properties. However, conventional computational methods often struggle with the formidable task of navigating the vast potential energy surface, especially in high-dimensional spaces with numerous local minima. Recent advancements in machine learning-driven surrogate models offer a promising avenue for alleviating this computational burden. In this study, we propose a novel approach that combines the strengths of universal machine learning potentials with a Bayesian approach using Gaussian processes. By using the machine learning potentials as priors for the Gaussian process, the Gaussian process has to learn only the difference between the machine learning potential and the target energy surface calculated for example by density functional theory. This turns out to improve the speed by which the global optimal structure is identified across diverse systems for a well-behaved machine learning potential. The approach is tested on periodic bulk materials, surface structures, and a cluster.
\end{abstract}

\maketitle

\section{Introduction}

The precise atomic configuration of a material significantly influences its mechanical, electronic, magnetic, and chemical properties. Determining this configuration at low temperatures computationally involves identifying the global minimum of the potential energy surface (PES). This optimization problem is notoriously challenging due to the vast configurational space and the presence of numerous meta-stable states. Conventional methods, such as basin hopping \cite{wales1997global}, particle swarm optimization, evolutionary algorithms \cite{vilhelmsen2014genetic}, and random searches \cite{pickard2011ab}, require extensive energy and force calculations, often making them impractical for large systems when utilizing computationally intensive methods like density functional theory (DFT).

\begin{figure}
    \centering
    \includegraphics[width=8.0cm]{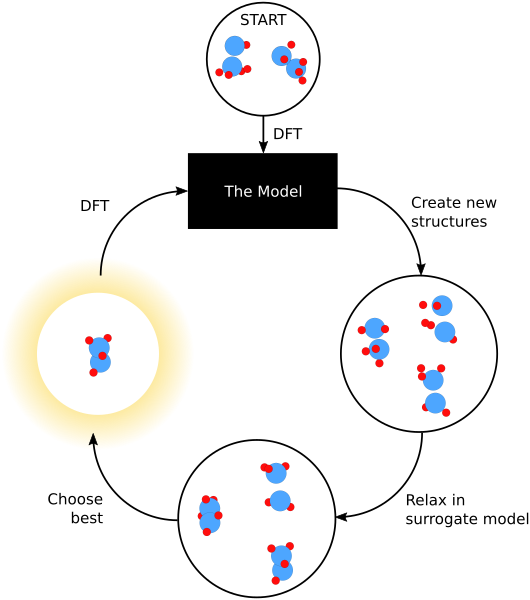}
    \caption{Illustration of the \texttt{GOFEE/BEACON} cycle. The model starts with an initial training set of a few random structures, which are evaluated with DFT and included in a database on which a Gaussian process surrogate model is trained.  The surrogate model is subsequently explored by minimizing the energy of a set of randomly picked initial structures. The obtained minima points on the surrogate surface are evaluated by means of an acquisition function, and the structure with the lowest value is selected. The energy and forces of this structure is calculated with DFT and the results are added to the database. The total number of cycles is set by the user.}
    \label{fig:beacon_cycle}
\end{figure}

The challenge of global optimization becomes even more pronounced in high-dimensional spaces where the PES landscape is riddled with local minima. Traditional methods require numerous DFT calculations, but recent advancements in machine-learned surrogate models offer promising alternatives that dramatically reduce the computational burden. Several approaches (BOSS \cite{todorovic2019bayesian}, GOFEE \cite{bisbo2020efficient}, \texttt{BEACON} \cite{kaapa2021global})
algorithms employ Gaussian processes to construct surrogate PES models, enabling efficient exploration through random searches and Bayesian optimization. These techniques have demonstrated orders of magnitude reduction in the number of necessary energy evaluations. They provide fast approximations of the PES by using training data obtained from initial DFT calculations and continuously expanded by means of an active leaning cycle as illustrated in Fig.~\ref{fig:beacon_cycle} for the \texttt{BEACON} method. As these algorithms start with minimal training data, often from a few random structures, they can be seen as ``cold-start'' approaches, which are initially rather uninformed compared to models trained on larger databases.

Concurrently, significant efforts have been dedicated to developing computational databases \cite{thygesen2016making,himanen2019data,saal2013materials,jain2013commentary,curtarolo2012aflow,draxl2019nomad,haastrup2018computational,gjerding2021recent,lyngby2022data,deng2023chgnet,merchant2023scaling} of atomic structures and basic material properties from high-throughput studies, which collectively contain results of millions of DFT calculations. These databases have enabled the rise of universal machine learning potentials (MLPs) \cite{chen2022universal,batatia2024foundation,choudhary2023unified,deng2023chgnet, merchant2023scaling,yang2024mattersim}, a promising tool in materials science. These potentials come pre-trained based on large datasets and can be applied out-of-the-box to different chemical systems across the periodic system. These potentials can predict the energy and forces with a surprising accuracy and several orders of magnitude faster than DFT making them ideal to use for the coarse energy landscape. However, while MLPs excel in capturing overarching trends, they may fall short in depicting the intricate nuances of potential energy surfaces.

In this paper we combine the comprehensive chemical knowledge of pre-trained MLPs with the Bayesian optimization approach of \texttt{BEACON}. By employing MLPs as a prior estimate of the energy and forces of a given structure, the Gaussian process focuses solely on learning the finer details of the energy landscape, while the MLP captures the general shape of the PES. We conduct comparative analyses across four different systems, spanning from periodic bulk materials to surface structures to a copper cluster, utilizing two distinct MLPs alongside the standard \texttt{BEACON} prior.

\begin{figure*}
    \centering
    \includegraphics[width=17.9cm]{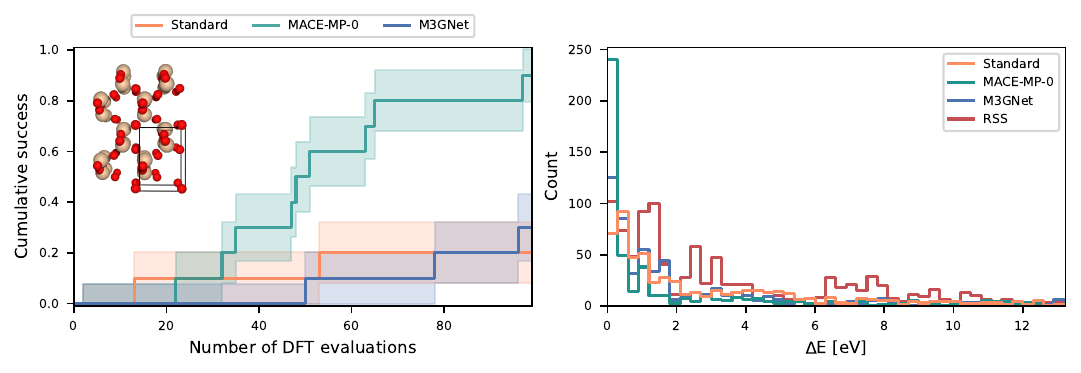}
    \caption{Left: Bulk SiO$_2$ success curve showing the cumulative success of finding the global minimum for each of the three priors. Shaded regions indicate the uncertainty based on Bayesian estimates (see Appendix).} Right: Histogram of all 1000 DFT energies of PES-relaxed structures obtained during the 10 runs. Also shown is 1000 random structure searches (RSS) where the corresponding structures are optimized using the MACE-MP-0 MLP.
    \label{fig:SiO2}
\end{figure*}

\section{Method}
\subsection{Surrogate energy surface}
The core surrogate model is based on the \texttt{GOFEE/BEACON} framework \cite{bisbo2020efficient, kaapa2021global} where the potential energy surface (PES) is modeled by a Gaussian process with gradients \cite{wu2017bayesian}. The energy and gradients, $\mu(\bm{x}) = \left( E(\bm{x}), -\bm{F}(\bm{x}) \right)$, of the given atomic configuration, $\bm{x}$, with a corresponding fingerprint $\rho(\bm{x})$ is given by
\begin{equation}\label{eq:predict}
    \mu(\bm{x}) = \mu_p(\bm{x}) + K \left( \rho(\bm{x}),P \right) C \left( P,P \right)^{-1} \left[ y - \mu_p(X) \right]
\end{equation}
where $C$ and $K$ are covariance matrices with and without regularization, $P=\rho(X)$ is a list of fingerprint vectors for the training data structures $X$ and $y$ is the training target energy and forces. $\mu_p(\bm{x})$ and $\mu_p(X)$ are the prior function for the given atomic configuration and the training data structures respectively. 
The associated uncertainty is given by

\begin{equation}
\resizebox{\linewidth}{!}{$\Sigma(\bm{x}) = [\Tilde{K}(\rho(\bm{x}),\rho(\bm{x}))-K(\rho(\bm{x}),P)C^{-1}(P,P)K(P,\rho(\bm{x})]^{1/2}$ }
\end{equation}

\noindent where $\Tilde{K}$ marks the covariance value of the predicted structure with itself.
From these equations we can formulate a \textit{lower confidence bound} acquisition function used to select the best candidate structure for the database given by \cite{bisbo2020efficient}

\begin{equation}
    A(\bm{x})=E(\bm{x})-2\Sigma(\bm{x}).
\end{equation}

As kernel we use a squared exponential kernel given by
\begin{equation}
k(\mathbf{x},\mathbf{x'})=\sigma^{2}\exp\Big(-\frac{D[\rho(\mathbf{x})-\rho(\mathbf{x'})]^{2}}{2l^{2}}\Big),
    \label{eq:fp_gaussiankernel}
\end{equation}
where $\sigma$ is a prefactor, $l$ is a length scale and $D$ marks the euclidean distance between two fingerprints. The hyperparameters $\sigma$ and $l$ are being updated after each iteration.

The fingerprint is a modified version of the invariant Valle-Oganov fingerprint \cite{valle2010crystal} which is split into radial parts, $\rho_{AB}^{R}$, and angular parts, $\rho_{ABC}^{\alpha}$, given by

\begin{equation}
\label{eq:fp_radial}
  \rho_{AB}^R(r; {\mathbf x}) = \sum_{\substack{i,j \\ i\neq j}} \frac{1}{r_{ij}^2}f_c(r_{ij}; R_c^R) \, e^{-|r-r_{ij}|^2/2\delta_R^2},
\end{equation}

\begin{equation}
\label{eq:fp_angular}
 \rho_{ABC}^\alpha(\theta; \mathbf x) = \sum_{\substack{ i,j,k \\ i\neq j \neq k }}  f_c(r_{ij}; R_c^\alpha) f_c(r_{jk}; R_c^\alpha) 
    e^{-|\theta-\theta_{ijk}|^2/2\delta_\alpha^2},
\end{equation}

\begin{equation}
\label{eq:cutoff}
f_c(r_{ij};R_c)=\begin{cases}
    \resizebox{.47\hsize}{!}{$1-(1+\gamma)\big( \frac{r_{ij}}{R_c} \big)^\gamma+\gamma\big(\frac{r_{ij}}{R_c}\big)^{1+\gamma} $}& \text{if } r_{ij}\leq R_c \\
    0& \text{if }r_{ij}>R_c
\end{cases},
\end{equation}
with $r$ and $\theta$ denoting atomic distances and angles respectively. The subscripts A, B, C signify different elements while $i$, $j$, $k$ signify atomic indices. $f_c(r_{ij};R_c)$ is a cutoff function going to zero at the radial and angular cutoff radii $R_c^R=4.25\Tilde{R}_{max}$ and $R_c^{\alpha}=2.25\Tilde{R}_{max}$ with $\Tilde{R}_{max}$ being the covalent radius of the largest element in the system. $\delta_R^{2}=0.4$Å, $\delta_{\alpha}^{2}=0.4$Å and $\gamma=2$ are constants. Discretization of $r$ and $\theta$ provides a vectorial fingerprint.
For more details on the specific definitions and equations we refer the reader to Ref.~\onlinecite{kaapa2021global}.

The \texttt{BEACON} algorithm's core principle is the iterative exploration of the PES through the following steps as seen in Fig. \ref{fig:beacon_cycle}: (1) An initial database is created, typically consisting of two DFT calculations based on random configurations. (2) A Gaussian process surrogate potential energy surface is constructed from the database. (3) The surrogate potential energy surface (PES) is explored through 40 energy minimizations, equaling the number of CPU cores available, starting from random configurations of atomic coordinates and the unit cell (if periodic). A total of 500 steps is taken in the minimizations. (4) The configurations are assessed using the \textit{lower confidence bound} acquisition function, and a DFT calculation for the configuration with the lowest acquisition function value is added to the database. The procedure then repeats from step (2). For an in depth description of the \texttt{BEACON} framework we again refer to Ref.~\onlinecite{kaapa2021global}.

All DFT calculations are performed using the Atomic Simulation Environment library \cite{larsen2017atomic} and the GPAW code \cite{enkovaara2010electronic, mortensen2024gpaw} with the PBE xc-functional \cite{perdew1996generalized}. For periodic structures we use a plane wave cut-off energy of \SI{800}{eV} and a \textit{k}-point density of \SI{6}{\AA}, while for the Cu$_{20}$ cluster we use a plane wave cut-off energy of \SI{400}{eV} and only the $\Gamma$ point is used for \textit{k}-point sampling.

The prior in a Gaussian process provides a flexible way to inform the model beyond the mere inclusion of data. This we are going to exploit further with the MLPs in the next section.

The standard prior energy, $E_p^0(\bm{x})$ used in BEACON is composed of a constant independent of atomic configuration plus a repulsive short ranged pair potential $E_\mathrm{rep}(\bm{x})$. (The prior appearing in Eq.~\ref{eq:predict} includes both the energy and the gradient $\mu^0_p(\bmx) = (E^0_p(\bmx), \nabla E^0_p(\bmx))$.)
The constant is set to the mean of all DFT energies calculated so far in the active learning cycle
\begin{equation}
E^0_p(\bmx) = \mathrm{Mean}(\{E_\mathrm{DFT}\}) + E_\mathrm{rep}(\bmx)
\end{equation}

The repulsive potential has the function of keeping atoms away from unphysical situations, where they are too close, and where DFT calculations would often fail \cite{bisbo2020efficient}. We use the form
\begin{equation}
        E_{\mathrm{rep}}(\bm{x})= 
\begin{dcases}
    \sum_{ij}  x_{ij}^{-12}- 12 (1 -x_{ij}) -1  ,& \text{if } x_{ij} \leq 1 \\
    0,              & \text{otherwise}
\end{dcases}
\end{equation}
where $x_{ij}=\frac{r_{ij}(\bm{x})}{\beta \left( \Tilde{R}_{i} + \Tilde{R}_{j} \right)}$, $ r_{ij}$ is the distance between atoms $i$ and $j$, and $\Tilde{R}_{i}$ and $\Tilde{R}_{j}$ denote their respective covalent radii. $\beta$ is a cutoff prefactor which determines when the repulsive force goes to zero. We use $\beta=0.7$.  The repulsive prior term is thus only active when the distance between atoms falls below a certain threshold.

\subsection{Machine Learning Interatomic Potential Prior}

The main idea behind the present paper is to substitute the rather uninformed standard prior used in BEACON with an MLP, so that the Gaussian process will have to learn only the difference between the MLP and the DFT calculations.

As a prior we investigate two MLPs: M3GNet \cite{chen2022universal} and MACE-MP-0 \cite{batatia2024foundation}, both trained on DFT relaxation snapshots from the Materials Project database \cite{jain2013commentary}. It's worth noting that the energies of materials in the Materials Project are computed using the VASP DFT code \cite{hafner1997vasp}, while our calculations employ the GPAW DFT code \cite{mortensen2024gpaw}. Due to the variance in reference energies between different DFT codes, a shift in predicted total energy occurs when comparing MLP-prior predicted energy to actual DFT-calculated energy. Nonetheless, the energy differences between distinct atomic configurations remain comparable. To address this energy shift, we introduce a modified constant term, $c$:
\begin{equation}
    E_p(\bm{x}) = E_{\mathrm{MLP}}(\bm{x}) + E_{\mathrm{rep}}(\bm{x}) + c,
    \label{eq:prior}
\end{equation}
which represents the mean energy difference between DFT calculations and MLP predictions:
\begin{equation}
c =\frac{1}{n_{\mathrm{steps}}} \sum_{i=1}^{n_{\mathrm{steps}}} \left( E_{\mathrm{DFT}}^{(i)} - E_{\mathrm{MLP}}^{(i)} \right)
\end{equation}
The constant is determined iteratively during the \texttt{BEACON} run as the mean energy difference between ongoing DFT calculations and MLP predictions.

We maintain the repulsive short ranged potential, $E_\mathrm{rep}$ in the prior because atomic configurations with small distances between the atoms pose challenges for the MLPs, as they are not designed to accurately describe these unrealistic structures. This challenge is particularly pronounced for MACE-MP-0 which may produce extremely negative or positive energy predictions, leading to erroneous identification of global minima or the creation of excessively high energy barriers within the surrogate energy landscape. We therefore also omit the $E_{\mathrm{MLP}}(\bm{x})$ in the prior when the minimum distance between the atoms are less than half the covalent radius of the smallest element in the atomic structure (i.e., $\text{min}(r_{ij})<0.5\Tilde{R}_{min}$). 
The repulsive potential together with the omission of the MLP ensures that atoms in close proximity experience a repulsive force, facilitating a smooth transition towards more physically realistic configurations where the MLP can accurately predict the energy.

\begin{figure*}
    \centering
    \includegraphics[width=17.9cm]{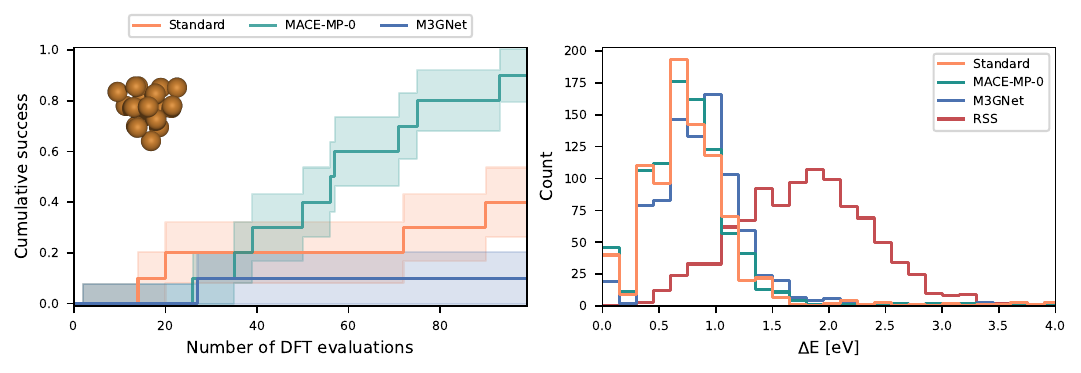}
    \caption{Left: Success curve for Cu$_{20}$ cluster. Right: Histogram of all DFT energies obtained during the runs.}
    \label{fig:Cu20}
\end{figure*}

\begin{figure*}
    \centering
    \includegraphics[width=17.9cm]{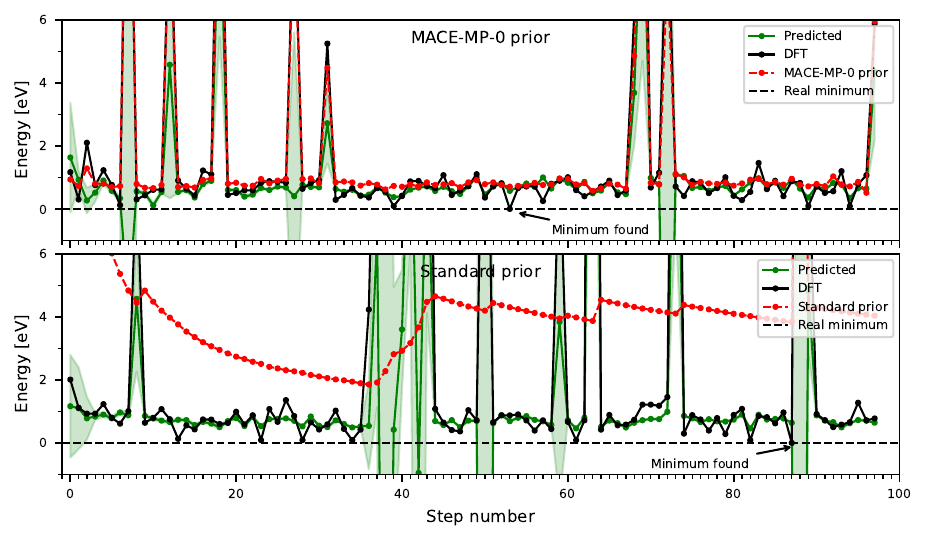}
    \caption{Progress of a single Cu$_{20}$ \texttt{BEACON} run showing the predicted, DFT-calculated, and prior-predicted energies for each step. The green shaded area shows the uncertainty as estimated by the Gaussian process. Top: using the MACE-MP-0 prior. Bottom: using the standard mean prior. The initial structures are not included in the plot.}
    \label{fig:Cu20_progress}
\end{figure*}

\section{Results}
\subsection{SiO$_2$}
The first system used to benchmark the MLP priors is bulk SiO$_2$ with 12 atoms in the unit cell. In this case, all degrees of freedom are allowed to be optimized, including both the atomic positions and the unit cell parameters.
To evaluate the performance of the different priors we use a so-called success curve, which shows the cumulative success of finding the global minimum of several individual runs as a function of the number of DFT evaluations performed. The cumulative curve increases by one step (with a step height of one divided by the number of individual runs) each time an individual run identifies the global minimum for the first time. If the global minimum is found in all runs, the curve will thus go to one after a certain number of DFT calculations.
In Fig.~\ref{fig:SiO2} (left), we present the success curve for bulk SiO$_2$. This curve illustrates the cumulative success rate of locating the global minimum across 10 individual runs, plotted against the number of DFT evaluations performed. We compare the success curves for three different priors: M3GNet, MACE-MP-0, and the standard \texttt{BEACON} prior, which is the mean of all DFT energies as well as the repulsion term, $\mu_{\mathrm{rep}}(\bm{x}) $, from Eq. \ref{eq:prior}. The \texttt{StructureMatcher} class from \texttt{pymatgen} \cite{ong2013python} is used to verify if the global minimum has been found. Our results indicate that the MACE-MP-0 prior significantly outperforms the standard prior, while M3GNet does not improve the success rate. We note that the unit cell is also optimized in this run which is the reason why the "Standard" prior is less likely to find the global minimum compared to the original paper \cite{kaapa2021global}, which used a fixed unit cell. 

At the end of each surrogate relaxation, the structure with the lowest acquisition function value is evaluated with DFT to determine its energy and forces. In a single \texttt{BEACON} run, the DFT energies of 100 candidate structures are calculated. The distribution of the DFT energies from the 1000 surrogate PES relaxations across the 10 runs for each prior is shown in Fig.~\ref{fig:SiO2} (right). Here we see that the MACE-MP-0 prior tends to yield lower energy structures compared to using M3GNet or the standard prior.

A critical question is whether the MACE-MP-0 MLP provides an advantage when used within the Bayesian \texttt{BEACON} framework or if it can independently find the global minimum. To investigate this, we performed a random structure search (RSS) where 1000 random structures were generated and optimized using the MACE-MP-0 potential, followed by DFT energy calculations. The energy distribution of these structures is also shown in Fig.~\ref{fig:SiO2} (right). The RSS approach produced higher energy structures compared to the \texttt{BEACON} framework, and the global minimum was found in only 0.9\% of the cases.

\subsection{Cu$_{20}$ cluster}
The MLPs utilized as priors are trained on databases of periodic bulk materials, and are thus expected to perform well in describing bulk systems. However, their efficacy on out-of-distribution structures, such as non-periodic clusters or surfaces, remains uncertain. MACE-MP-0 has previously demonstrated notable out-of-distribution performance. To evaluate its effectiveness within the \texttt{BEACON} framework, we conducted \texttt{BEACON} simulations on a cluster of 20 copper atoms. This cluster has two distinct low-energy structures: the global minimum and a local minimum 0.1 eV higher. Figure \ref{fig:Cu20} presents the success curve for the cluster, showing that MACE-MP-0 surpasses the standard prior in performance. Conversely, the use of the M3GNet prior appears to hinder the algorithm's efficiency compared to the standard prior. For this cluster, the RSS method fails to locate the global minimum or the second lowest structure in any of the 1000 MACE-MP-0 MLP relaxations. Generally, the distribution of RSS energies is markedly shifted towards higher values compared to all the \texttt{BEACON} runs (refer to Fig.~\ref{fig:Cu20}, right).

Let's examine the progress of a single run with the MACE-MP-0 prior and the standard prior, as illustrated in Fig.~\ref{fig:Cu20_progress}. This figure displays the predicted, DFT-calculated, and prior-predicted energies for each step in the \texttt{BEACON} cycle. The MACE-MP-0 prior effectively captures the overall quantitative energy landscape, accurately distinguishing between high and low energy structures. However, it struggles with the finer details of the PES near the lowest energy structures. For instance, the MLP fails to predict the energy dip at step 39 and does not identify the global minimum at step 53.

\begin{figure*}
    \centering
    \includegraphics[width=17.9cm]{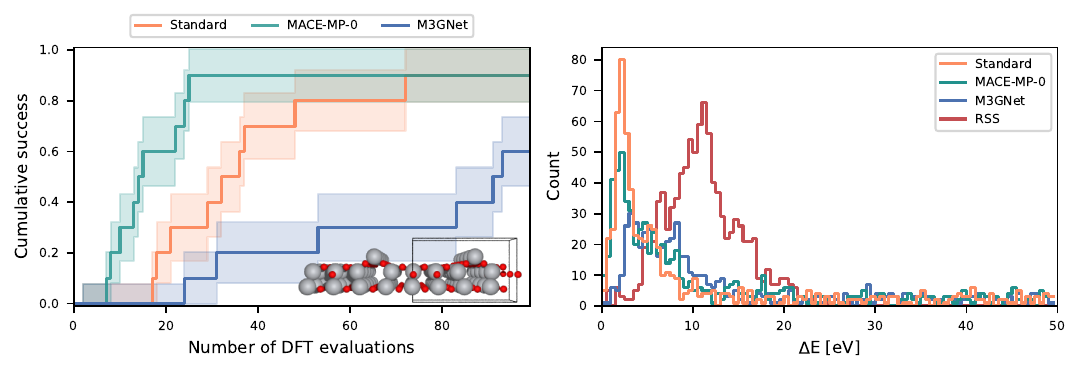}
    \caption{Left: Success curve for anatase TiO$_2$(001)-(1 × 4) surface reconstruction. Right: Histogram of all DFT energies obtained during the runs.}
    \label{fig:TiO2}
\end{figure*}

\subsection{Anatase TiO$_2$(001)-(1 × 4) surface reconstruction}
Another well-studied benchmark system is the anatase TiO$_2$(001)-(1 × 4) surface reconstruction \cite{lazzeri2001stress,bisbo2022global}, characterized by periodic rows of protruding Titanium atoms (see Fig.~\ref{fig:TiO2}). In our study, we consider a configuration of 27 atoms, with the bottom 12 atoms fixed in the bulk structure while the top 15 atoms are optimized, keeping the unit cell fixed. The resulting success curve is displayed in Fig.~\ref{fig:TiO2} (left), showing that the MACE-MP-0 prior again outperforms the standard prior. Similar to the case of Cu${20}$, the M3GNet prior seems to impede the algorithm's ability to find the minimum, leading to poorer performance.

Remarkably, \texttt{BEACON} with the MACE-MP-0 prior identifies the global minimum with only approximately 20 single-point calculations, a significant improvement compared to the study by Bisbo and Hammer \cite{bisbo2022global}, where their GOFEE algorithm required around 300 single-point calculations. This performance enhancement of \texttt{BEACON} can be largely attributed to its generalization to include training on both forces and energies, whereas GOFEE only utilizes energies. This is further supported by the fact that standard \texttt{BEACON} also successfully identifies the global minimum in approximately 40 single point calculations.

Examining all the DFT energies during the runs in Fig.~\ref{fig:TiO2} (right) and comparing them with the RSS results, we observe that RSS tends to find higher energy structures compared to \texttt{BEACON}. RSS identifies the global minimum in only 9 out of the 1000 MACE-MP-0 optimized structures.

\section{Assessing the ML potentials}
\begin{figure*}
    \centering
    \includegraphics[width=17.9cm]{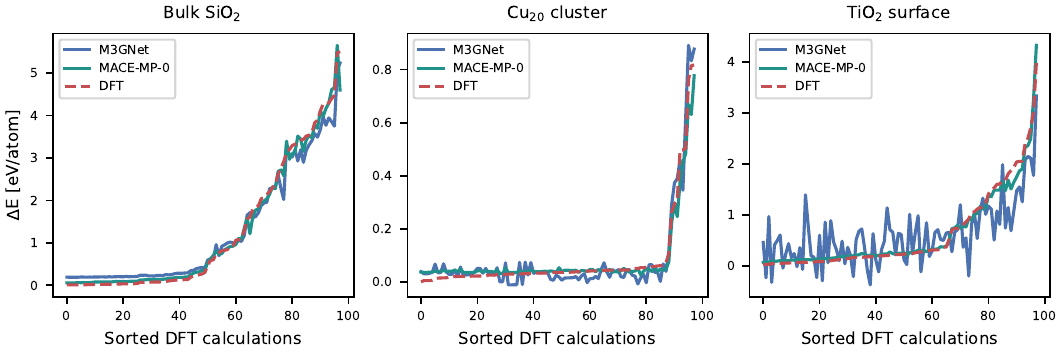}
    \caption{ 
    Validation of the MLPs for a single BEACON run across the three materials. Energies of the 100 DFT-validated structures are predicted using the two MLP models, with an offset applied to each MLP by the mean energy difference relative to DFT. $\Delta$E represents the energy difference from the true DFT energy of the ground-state structures. All energies are sorted by ascending DFT energy. MACE-MP-0 is seen to provide much better predictions than M3GNet.}
    \label{fig:energies}
\end{figure*}

In all three test systems, the MACE-MP-0 prior consistently improves performance, whereas M3GNet either matches the performance of the standard prior (as seen for SiO$_2$) or underperforms (as with the Cu$_{20}$ cluster and the TiO$_2$ surface). To investigate these differences, we analyze a single BEACON run to examine each MLP’s modeling of the PES. Specifically, we use the 100 DFT-validated structures generated in a single run, and predicts their energies with both MLPs. Figure ~\ref{fig:energies} shows the (offset-corrected) predicted energies of the two priors alongside the true DFT energies for these structures. The energies are ordered by ascending DFT energy, revealing trends across energy levels.

For all three materials, MACE-MP-0 accurately captures the PES, both near the global minimum and at higher energy structures. In contrast, M3GNet fails to reliably represent the PES for the cluster and surface, showing substantial fluctuations in predicted energies near the global minimum. These fluctuations make it more challenging for the Gaussian process to learn the difference between the prior and the correct energy, while the smoother and more consistent MACE-MP-0 prior facilitates this learning. This difference accounts for M3GNet’s negative impact on \texttt{BEACON} performance in the Cu$_{20}$ cluster and TiO$_2$ surface cases.

It is worth noting that both M3GNet and MACE-MP-0 were trained on bulk structures, which makes their application to clusters and surfaces out-of-distribution. However, MACE-MP-0 has demonstrated strong out-of-distribution performance in previous studies \cite{batatia2024foundation}, which aligns with our findings here.

\section{Conclusion}
In this study, we present a novel approach to the optimization of atomic structures by integrating the strengths of universal machine learning potentials with the Bayesian optimization framework provided by \texttt{BEACON}. Our method leverages pre-trained MLPs as a prior to capture the general features of the potential energy surface, allowing the Gaussian process within the \texttt{BEACON} framework to focus on the intricate details that define the global minima. We have shown that the MACE-MP-0 prior improves the \texttt{BEACON} framework resulting in a higher success rate of finding the global minimum across various systems, including periodic bulk materials, surface structures, and a copper cluster.

The GOFEE/BEACON framework was recently extended with the aim of removing energy barriers on the surrogate PES. For example, \texttt{ICE-BEACON} \cite{kaappa2021atomic} allows for interpolation between chemical elements, and \texttt{ghost-BEACON} \cite{larsen2023machine} introduces atoms of fractional existence as candidate sites for real atoms. These extensions require machine-learning predictions for unphysical systems with for example atoms, which are partly gold and partly copper. The Gaussian process is able to provide this with a simple generalization of the fingerprint in Eqs.~(\ref{eq:fp_radial})-(\ref{eq:fp_angular}) \cite{kaappa2021atomic}. However, it is presently not clear if the MLP priors can somehow be generalized to systems with fractional chemical identity or existence, so that the improvements seen in this work can be combined with the speed-ups obtained with the ICE and ghost approaches.

\section{Acknowledgements}
We acknowledge VILLUM Fonden  for grant no. 37789 as well as VILLUM Center for Science of Sustainable Fuels and Chemicals, which is funded by the VILLUM Fonden research grant 9455.

\section{Competing interests}
The authors declare no competing interests.

\section{Data Availability}
The code used is available at \\ \url{https://gitlab.com/gpatom/ase-gpatom/-/tree/ML_prior}.

\section{Appendix}

\subsection{Structure generation}
In step (2) of the \texttt{BEACON} cycle, random structures are generated by placing atoms randomly within a box or a randomly generated unit cell (if periodic conditions apply). The box size and unit cell sizes are determined by the total volume of atomic spheres, $V_{cr}$, where the radii correspond to the covalent atomic radii of the elements.

For periodic SiO$_2$, the unit cell volume is set to be four to ten times $V_{cr}$. In contrast, for the Cu$_{20}$ cluster, a box with a volume 1.5 to 3.5 times $V_{cr}$ is used while the unit cell length is fixed to 22 Å. For the TiO$_2$ surface, the randomly placed atoms are confined to a box with a height of 2.5 Å above the fixed atom layer. 

After the initial random placement, the atoms are repelled until all atom centers are at least 0.95 times the summed covalent atomic radii apart from each other.

\subsection{MLP details}
The M3GNet model is loaded through the \textit{matgl} Python library, with the specific model "M3GNet-MP-2021.2.8-PES". For the MACE-MP-0 model, we employed the "medium" version. Both MLPs are employed in their pretrained state, with no further fine-tuning applied.

\subsection{Success curve uncertainty}
The success curve is modeled as a Beta distribution as in Ref.~\onlinecite{larsen2023machine}. Here, the success curve is represented as $ n + m$ independent attempts to find the global optimum, where $n$ and $m$ denote the numbers of successful and unsuccessful attempts, respectively. Applying Bayes' theorem with a uniform prior, the posterior probability of success $ p_s $ follows a Beta distribution, given by $B(p_s \,|\, \alpha = n + 1, \, \beta = m + 1) $. We use the mode of this distribution, $\text{mode}(p_s) = {n}/(n + m)$, as the value of the success curve. For the uncertainty, we adopt the square root of the variance of the distribution,

\begin{equation}
    \sqrt{\mathrm{Var}\left( p_s \right)} = \sqrt{\frac{(n+1)(m+1)}{(n+m+2)^2 (n+m+3)}}
\end{equation}

\subsection{Computational cost}
All calculations were performed on two Intel Xeon Gold 6148 CPUs, totaling 40 cores. GPAW DFT calculations were fully parallelized across all 40 cores, while surrogate relaxations were run on a single core, with 40 independent relaxations processed in parallel. We use a maximum of 500 single-point energy/force evaluations in the surrogate relaxations.

The computational time for each stage of the BEACON process varies depending on system characteristics and hardware configuration. These stages include DFT calculations, Gaussian process training, fingerprint calculation, MLP-based prior prediction, and Gaussian process application, each requiring different amounts of time relative to the others.

For the Cu$_{20}$ cluster, a single DFT calculation using 40 cores requires approximately 1 minute, while 500 single-point evaluations using MACE-MP-0 on a single core also take around 1 minute.

For the TiO$_{2}$ surface, DFT calculations take about 8 minutes, and 500 single-point evaluations with MACE-MP-0 using a single core require approximately 2 minutes.

For bulk SiO$_{2}$, DFT calculations take slightly over 1 minute, with 500 single-point evaluations using MACE-MP-0 on a single core also taking slightly over 1 minute.

The M3GNet MLP is approximately 5–10 times faster than MACE-MP-0, while the time of the standard prior is negligible due to its analytical nature.

\bibliography{refs}

\end{document}